%% file: TimeDomainAGN.tex
\RequirePackage{ifpdf}
\documentclass{PoS}
\usepackage[authoryear,square]{natbib}
\bibpunct{(}{)}{;}{a}{}{,}

\title{Time domain studies of Active Galactic Nuclei with the Square Kilometre Array}

\ShortTitle{Time domain studies of AGN}

\author{Hayley Bignall,$^1$ 
        \speaker{Steve Croft},$^{2,3}$
        Talvikki Hovatta,$^{4,5}$ 
        Jun Yi Koay,$^6$ Joseph Lazio,$^7$ Jean-Pierre Macquart$^{1,8}$
        and Cormac Reynolds$^1$\\
\llap{$^1$}ICRAR/Curtin University\\
\llap{$^2$}University of California, Berkeley\\
\llap{$^3$}Eureka Scientific\\
\llap{$^4$}California Institute of Technology\\
\llap{$^5$}Aalto University, Mets\"ahovi Radio Observatory\\
\llap{$^6$}DARK, Niels Bohr Institute, University of Copenhagen\\
\llap{$^7$}Jet Propulsion Laboratory, California Institute of Technology\\
\llap{$^8$}ARC Centre of Excellence for All-Sky Astrophysics (CAASTRO)\\
E-mail:  \email{H.Bignall@curtin.edu.au}, \email{scroft@astro.berkeley.edu}}

\abstract{
Variability of radio-emitting active galactic nuclei can be used to probe both intrinsic variations arising from shocks, flares, and other changes in emission from regions surrounding the central supermassive black hole, as well as extrinsic variations due to scattering by structures in our own Galaxy. Such interstellar scattering also probes the structure of the emitting regions, with microarcsecond resolution. Current studies have necessarily been limited to either small numbers of objects monitored over long periods of time, or large numbers of objects but with poor time sampling. The dramatic increase in survey speed engendered by the Square Kilometre Array will enable precision synoptic monitoring studies of hundreds of thousands of sources with a cadence of days or less. Statistics of variability, in particular concurrent observations at multiple radio frequencies and in other bands of the electromagnetic spectrum, will probe accretion physics over a wide range of AGN classes, luminosities, and orientations, as well as enabling a detailed understanding of the structures responsible for radio wave scattering in the Galactic interstellar medium.
          }

\FullConference{
Advancing Astrophysics with the Square Kilometre Array\\
June 8-13, 2014\\
Giardini Naxos, Italy}

\DeclareRobustCommand{\ion}[2]{%
\relax\ifmmode
\ifx\testbx\f@series
{\mathbf{#1\,\mathsc{#2}}}\else
{\mathrm{#1\,\mathsc{#2}}}\fi
\else\textup{#1\,{\mdseries\textsc{#2}}}%
\fi}

\begin{document}
\makeatletter
\setbox\@firstaubox\hbox{\small Hayley Bignall}
\makeatother

\input{sec-intro.tex}

\input{sec-science.tex}
In \S\ref{sec:intrinsic} below we discuss direct observations of intrinsic AGN variability, with reference to many recent studies. In \S\ref{sec:scat} we discuss variability due to scattering in the interstellar medium, including results from the extensive MASIV VLA Survey and follow-up observations, and still-mysterious ESEs. Monitoring of ISS probes intrinsic source changes, as well as small-scale structures in the ISM. In \S\ref{sec:classification} we discuss methods for disentangling variations due to intrinsic and extrinsic mechanisms.

\input{subsec-science-intrinsic.tex}

\input{subsec-science-scattering.tex}

\input{subsec-classification.tex}

\input{sec-SKA-capabilities.tex}

\section{Summary}
Variability of AGN provides a powerful probe of the physics of accretion and outflow, as well as the dynamics and structure of our own Galaxy's interstellar medium. The huge leap in survey speed engendered by SKA will enable a precision synoptic view of variability for hundreds of thousands of AGN, an archive of variability statistics for comparison to LSST and other large surveys, and a huge expansion into unexplored parameter space as the study of radio AGN truly enters the wide-field time-domain era. 
These data will allow us to move from the phenomenology and statistics of radio variability to understanding the physical mechanisms responsible for the large range of observed variability behaviour. 
Key capabilities for this work are broad frequency coverage---it is particularly desirable to include the 4.6-13.8~GHz Band 5 on SKA-MID; flexible, independent sub-arrays which can cover different frequency bands; and high angular resolution capability.

\acknowledgments
SDC acknowledges support from the Theodore Dunham, Jr. Grant of the F.A.R.\ and from the NSF via AST-1009421. TH acknowledges support from the Academy of Finland project number 267324. JYK acknowledges support by a research grant (VKR023371) from VILLUM FONDEN. The Dark Cosmology Centre (DARK) is funded by the Danish National Research Foundation. Part of this research was carried out at the Jet Propulsion Laboratory, California Institute of Technology, under a contract with the National Aeronautics and Space Administration. We thank the reviewers for helpful feedback to improve the manuscript.

\bibliographystyle{truncapj}
\bibliography{TimeDomainAGN}

\end{document}

%% file: sec-intro.tex
\section{Introduction}
The limited survey speed of existing radio telescopes means that active galactic nucleus (AGN) variability has been studied in detail only for $\sim 10^3$ relatively bright objects, or in only a few epochs for larger numbers of sources down to lower flux density limits.
However, AGN dominate the radio sky at flux densities above a milliJansky \citep{seymour:08}, and synoptic surveys with SKA Phase 1 will provide radio variability information with a cadence of days or less for hundreds of thousands of AGN. 

The characteristic timescales of the inner regions of the AGN central engine are accessible on human timescales, from hours to days to years, and enable processes occuring close to the supermassive black hole (SMBH) to be probed. Understanding variability on these timescales, and correlating variability over a range of radio frequencies with that seen across the electromagnetic spectrum, can provide a powerful probe of the connection between disks and jets, inflow and outflow, turbulence and shocks, and activity and quiescence. 

In addition to source-intrinsic processes, propagation effects also play an important role in observed variability at centimetre wavelengths. Interstellar scintillation (ISS) in the strong refractive and weak scattering regimes produces apparent intensity variations for sufficiently compact sources on timescales of typically days, but sometimes as short as minutes to hours. ISS gives access to microarcsecond ($\mu$as) scales \citep{Lovell08}, allowing the study of changes in intrinsic source structure with resolution beyond that achievable with VLBI. Extreme scattering events (ESEs) are transient events in which sources undergo large, week- to months-long time-symmetric flux density excursions, due to refractive lensing potentially by discrete structures in the interstellar medium (ISM). While these effects have been known for decades, several significant questions remain, both with respect to the characteristics of sources that vary and the interstellar structures that produce the variation.

In this Chapter we discuss the contribution that SKA will make to various outstanding questions in time domain studies of AGN, through observations of both intrinsic and extrinsic variability. Although a broad range of specific science outcomes is addressed, similar or identical survey strategies are involved, and disentangling intrinsic and extrinsic causes of observed variability will be important in order to achieve the desired goals: increased understanding of the physics both of AGN and of the intervening media.

%% file: sec-science.tex
\section{Science from time domain studies of AGN: state of the art}

%% file: subsec-science-intrinsic.tex
\subsection{AGN physics from intrinsic variability\label{sec:intrinsic}
}

Variability of both total intensity and polarization is a common characteristic of AGN, and occurs with multiple characteristic timescales in individual objects. Variations between different bands of the electromagnetic spectrum are often related, with early emission at short wavelengths---X-rays or $\gamma$-rays---eventually propagating to longer wavelengths, including radio, and longer timescales. Emission at different wavelengths probes a range of processes occuring in different regions of the AGN central engine, and a panchromatic view can help to answer questions of how relativistic AGN jets are launched, and how inflow and outflow are coupled. In some cases \citep[e.g.][]{thyagarajan:11} variable radio emission is the only indicator of AGN activity, and can provide a powerful probe of buried AGN which show no evidence of an accreting black hole at other wavelengths. 

AGN intrinsic variability exhibits both red and white noise components, as well as quasi-periodic oscillations in some cases \citep{king:13,wiita:11}, corresponding to variations
in accretion rate, flares and shocks in disks and jets, transitions
between high and low states \citep{krauss:13}, changes in Doppler boosting and jet precession, and other processes taking place near
the black hole. Long-term monitoring observations are required to sample the full range of timescales, to compare to observations at other wavelengths, and to catch rare events such as large flares and outbursts. The correlation delay between X-ray and radio can be hundreds of days, so long duration campaigns (at least several years) are required for adequate sampling. Day to year timescales for AGN variability correspond to the light crossing, viscous, and infall timescales of the black hole and accretion disk. Several ongoing programs \citep[e.g.][]{hughes:92,hovatta:07,lister:09,richards:11} have monitored a few hundred sources for significant periods of time. A few other surveys \citep[e.g.][]{thyagarajan:11} have obtained data over a handful of epochs for large numbers of sources. But until now, the primary limitation for this exercise has been the absence of dedicated radio instruments with the combination of field of view and sensitivity to partner with high energy facilities. SKA will open new parameter space by providing precision flux measurements as a function of time for almost all of the hundreds of thousands of compact sources typical of previous large-area single-epoch surveys. By providing high-cadence synoptic monitoring without the potential biases of small sample selections, SKA will produce an archive of variability information for a huge range of AGN.

The launch of the {\em Fermi} Space Telescope in 2008 revolutionized gamma-ray studies of AGN by providing the first instrument that monitors the entire gamma-ray sky every three hours \citep{atwood:09}. It has become possible to do population studies of gamma-ray loud, and quiet, AGN and compare the gamma-ray emission to their radio properties. For example, gamma-ray detected AGN have larger variability amplitudes in radio \citep{richards:11}, faster apparent jet speeds \citep{lister:09b}, and higher Doppler beaming factors \citep{savolainen:10} than non-detected objects. 
Some studies have also reported that radio and gamma-ray luminosities are correlated \citep{ackermann:11, nieppola:11}. All of the above imply a connection between radio and gamma-ray variations in the sources. The next generation very high energy Cherenkov Telescope Array (CTA), in combination with SKA, will in turn open the possibility to study the connection between TeV and radio emission in a large number of sources, probing the acceleration of particles up to the highest energies \citep{giroletti}.

Many existing surveys have focused on blazars, where relativistic beaming and the compact sizes of emitting regions produce more variability than in sources oriented further from the line of sight. The wide field monitoring capabilities of SKA, however, will enable lightcurves from all classes of AGN to be studied, refining unification schemes and our understanding of the geometry of the emitting regions. One caveat is that most, but by no means all \citep[e.g.][]{Kedziora01,tanami,stevens:12}, of the existing monitoring surveys were undertaken in the northern hemisphere, so not all of the sources from these surveys will be accessible for continued monitoring with SKA.

\begin{figure}
\centering 
\includegraphics[angle=0,width=.5\textwidth]{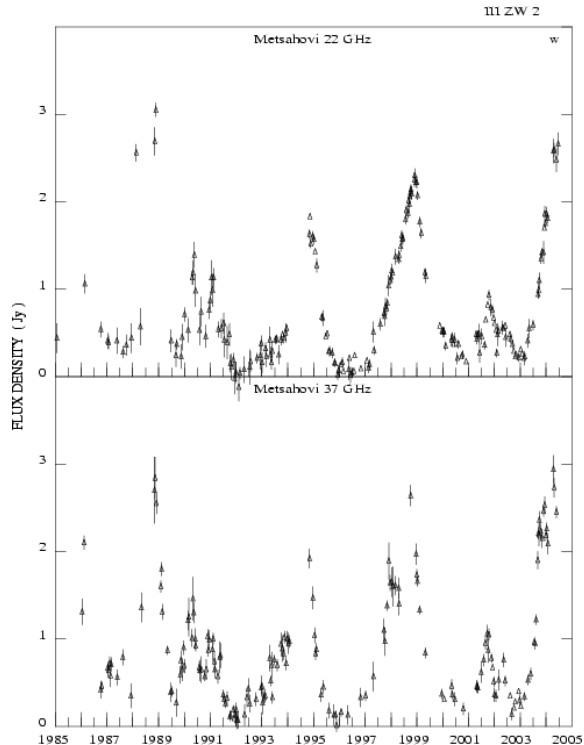}
\caption[]{Lightcurve of III\,Zw\,2 at 22 and 37\,GHz from the Mets\"{a}hovi Radio Observatory, showing flaring activity suggestive of acceleration of jets from non-relativistic to relativistic speeds during outbursts. Credit: \citeauthor{terasranta:05}, A\&A, 440, 409, 2005, reproduced with permission~\textcopyright~ESO.}
\label{fig:iiizw2}
\end{figure}

The details of accretion, and its connection to jet launching in radio-loud sources, as well as a wide range of other AGN phenomena, are still not well understood. 
Radio flares may occur at time intervals corresponding to the formation of shocks in the jets, which could be due to interaction with the ISM or torus \citep{brunthaler:05}, or to changes in accretion rate propagating to changes in the jet \citep{chatterjee:11}. 
The most commonly used models for flares in blazars are so called shock-in-jet models where the variations are caused by shocks propagating down the jet and emitting synchrotron radiation \citep{marscher85, hughes85}. Multifrequency radio observations can be used to model the shock propagation and jet properties in detail, giving insights into the jet geometry and magnetic field \citep{fromm11}.
Radio flares seen in III\,Zw\,2 (Fig.~\ref{fig:iiizw2}) and Mrk 348 suggest that jets in Seyferts can accelerate from non-relativistic to relativistic speeds during an outburst, but larger samples are required to confirm this \citep{mundell:09}. 
Interestingly, narrow-line Seyfert I galaxies are an unexpected group of gamma-ray emitting AGN. Fermi has detected a handful of NLS1 galaxies, confirming that they indeed can also emit high energy electrons, and have prominent relativistic jets \citep{abdo09}.
Modeling these variations can probe jet acceleration mechanisms and central engine physics \citep{kud:11}. Understanding jet production is also critical in studying galaxy formation, since jets play a key role in regulating galaxy growth \citep{hopkins:10}.

Another promising avenue for understanding accretion and jet launching is the study of tidal disruption events (TDEs), posited to result from stars that are shredded after passing close to SMBH event horizons. The sudden increase in accretion rate results in SMBHs transitioning from a quiescent state to a short-lived, typically months-long, AGN-like phase. In some cases radio emission is thought to result from the formation of a jet, which interacts with the dense ISM of the galaxy \citep{metzger:11,zauderer:11,burrows:11}, and produces a radio afterglow that peaks around a year after the star is tidally disrupted and lasts for around a year. A large sample of TDEs would allow us to address the unanswered question of how jets evolve when accretion rates undergo a sudden dramatic increase. By looking for common features in TDEs that are radio loud versus radio quiet, their role in AGN radio loudness in general could be constrained \citep{svv:11}. TDEs are expected to occur in an average galaxy about once every $10^5$\,yr, so large surveys such as those that will be undertaken by SKA are required to see them in significant numbers. Further discussion of the role of SKA in finding TDEs can be found elsewhere in this volume \citep{donnarumma}.

A recent study \citep{macleod:10} found a dependence of AGN variability on black hole mass, luminosity, and Eddington ratio, as well as larger variability amplitudes for radio-loud objects than for radio-quiet. However, the relationship between variability and the properties of the host galaxy and black hole, and the accretion dynamics, are by no means clear \citep{zuo:12}. Recently, it was shown that the jet magnetic field and accretion disk luminosity are tighly correlated in blazars \citep{zamaninasab14}. This result was obtained by comparing black hole mass estimates to the jet magnetic flux, calculated using radio interferometric observations. This result shows that it is possible to obtain information about the detailed physics near the black hole, by relating observations further out in the jet to the smallest scales near the black hole.

VLBI observations provide direct evidence for relativistic outflows from AGN, and have been used to monitor the motion of jet components on parsec scales \citep[e.g.][]{lister13}. It is possible to determine jet compositions, magnetic fields, variations in jet orientation, ages and densities of components, interplay between the jet and ambient medium, as well as other physical conditions in the jet and close to the black hole by comparing parsec-scale variations with spectral energy distributions, polarization and Faraday rotation, as well as observations at other wavelengths and relativistic MHD simulations \citep{hovatta12,hovatta14}. VLBI components are by their very nature intrinsically variable \citep{lister:09}, but are also subject to variability due to interstellar scattering (\S\ref{sec:scat}). Flares and superluminal blobs may be connected to changes in magnetic field on timescales of years, or to current-driven instabilities, tangled magnetic fields, or reconnection events, and SKA can help probe this \citep{agudo}.

\begin{figure}
\centering 
\includegraphics[angle=0,width=.8\textwidth]{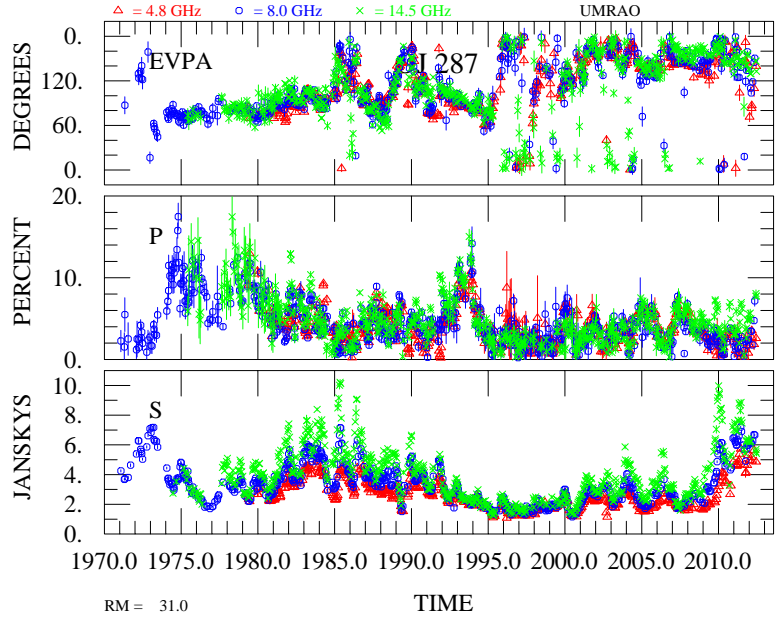}
\caption[]{Centimetre-band lightcurves of BL Lac object OJ\,287 from the University of Michigan Radio Telescope. Optical lightcurves of this source, including archival data dating back over 100 years, show a 12 year periodicity interpreted as the orbital period of a secondary SMBH intercepting the accretion disk of the primary. The radio lightcurves shown here (from \citealt{aller14}, reproduced by permission of the AAS) show flaring activity on shorter timescales. From top to bottom, the panels show total flux density, fractional linear polarization, and polarization electric vector position angle.}
\label{fig:oj287}
\end{figure}

AGN activity appears to be enhanced during galaxy mergers \citep{manzer:14} and in overdense environments \citep{croft:07}. However, the relationship between mergers and the triggering of various kinds of AGN, including those selected at different frequencies, is still not clear \citep{kocevski:12,karouzos:14}. The details of what happens to SMBHs after mergers are also not yet well understood. In some instances, AGN pairs (or multiple AGNs) with an offset in position and / or velocity are seen in galaxies at various stages of mergers \citep[e.g.][]{komossa:03,comerford:11}. With large area, sensitive, high resolution SKA surveys, many more dual AGN (including binary AGN, where the AGN are clearly orbiting each other, e.g. \citealt{rodriguez:06}; \citealt{deane:14}) or systems containing an AGN and a quiescent SMBH \citep[e.g.][]{comerford:14}, should be detected at various stages in their evolution. In at least one case known to date, the blazar OJ\,287, variability of optical emission is proposed to arise as a secondary black hole plunges through the accretion disk of the primary \citep{valtonen:08}. The optical lightcurve shows periodic outbursts every 12 years, corresponding to the proposed orbital period, and radio monitoring (Fig.~\ref{fig:oj287}) shows evidence of long term cycles of radio activity as well as periodic variability associated with flares and jet features \citep{hughes98}.

For closer binaries, the ``final parsec problem''---the ineffectiveness of dynamical friction with stars near SMBHs in shrinking their separation to $<1$\,pc \citep{finalparsec}---remains unsolved, and likely requires a more detailed understanding of accretion disk dynamics. Periodic variability due to jet precession, a signature of gravitational wave dominated inspiral of SMBHs during the final stage of mergers \citep{haiman:09}, can be used to search for merging systems containing a radio-loud AGN. A detection of this signal, which should be within reach of SKA and its precursors \citep{osh:11}, could be used to probe the physics of viscous orbital decay as a function of black hole mass, and confirm that inspiral is driven by gravitational wave emission.

By probing the inner regions of AGN central engines, studies of AGN intrinsic variability with SKA will inform two Key Science Projects: ``Strong Field Tests of Gravity Using Pulsars and Black Holes'', and through the connections between infall and outflow in AGN, and feedback on surrounding gas, ``Galaxy Evolution, Cosmology, and Dark Energy''.

%% file: subsec-science-scattering.tex
\subsection{Science from interstellar scattering\label{sec:scat}}
%

Radio wave propagation effects and implications for SKA, applying to studies of both pulsars and AGN, were discussed in some detail by \citet{Lazio04}. Here we provide an updated overview specifically related to AGN variability due to interstellar scattering, and outline the currently outstanding questions that can be addressed with SKA.

\subsubsection{Interstellar scintillation}
Over the past two decades, evidence has accumulated to demonstrate that rapid---intra-day and inter-day---flux density variations of AGN at centimetre wavelengths are predominantly caused by ISS rather than intrinsic variability.
This evidence includes the frequency dependence of the observed variability \citep{Kedziora97}, measurement of variability pattern arrival time delays of several minutes between widely separated telescopes for the most rapid variations \citep{Jauncey00,Dennett02,Bignall06}, annual cycles in the characteristic timescale of variability \citep[e.g.][]{Dennett01,Rickett01,Jauncey01,Qian01,Bignall03} and a strong correlation between cm-wavelength intraday variability (IDV) and Galactic H$\alpha$ emission measure along the line-of-sight, showing that this IDV is related to the ionized Galactic ISM \citep{Lovell08,Koay11}.

Observed ISS implies the presence of high brightness temperature compact components, no larger than tens of $\mu$as in angular size \citep[e.g.][]{Lovell08}. ISS, with an adequate model of the scattering parameters, can thus be used to estimate source angular size and compact fraction, as well as investigate frequency- and polarization-dependent sub-structure, at resolutions beyond those achievable with ground-based VLBI. Angular scales of this order are presently being probed directly on baselines to the RadioAstron Space Radio Telescope \citep{Kardashev13}, but the sensitivity is limited to bright sources. Studies of ISS of fainter sources over wide bandwidths probe a fundamentally different, much weaker, regime of AGN jet power.

The MASIV VLA Survey at 5\,GHz \citep{Lovell03,Lovell08} found that more than half of all compact, flat-spectrum sources $>100$\,mJy exhibited rms variations $\geq 2$\% on timescales $\leq 3$ days due to ISS, with increased fractional variability towards lower flux densities. 
This trend is as expected for brightness temperature-limited sources, where angular size is expected to scale as $\sqrt{S}$ assuming no change in the Doppler boosting factor. The majority of MASIV sources have firm spectroscopic identifications---all AGN, with almost 80\% being flat-spectrum radio quasars and most of the remaining fraction BL Lac objects \citep{Pursimo13}. 

A suppression of ISS towards higher redshift sources \citep{Lovell08} has been proposed as a possible indicator of angular broadening due to scattering in the turbulent intergalactic medium (IGM), making the high redshift sources appear too large to scintillate through the Galactic ISM. However, a careful analysis of the MASIV Survey and follow-up VLA data at 5 and 8 GHz found no conclusive evidence for IGM scatter-broadening of the high redshift MASIV sources. After accounting for a steepening with redshift of average spectral index between 5 and 8 GHz, the decrease in amplitude of ISS with redshift was found to be no longer significantly in excess of that expected if angular size $\theta \propto \sqrt{1+z}$, as expected for a brightness temperature and flux limited sample of sources \citep{Koay12}.

The statistics of short timescale radio variability for a larger and much fainter source sample, obtainable with SKA1, will allow tighter limits to be placed on extragalactic scatter-broadening through the suppression of ISS. As lower flux density AGN cores can have smaller intrinsic source sizes, they are more likely to be dominated by scatter-broadening. Observations of large numbers of faint sources in fields at high Galactic latitudes would provide a means of determining the ISM contribution to scattering; source redshifts are also required in order to measure, or obtain limits on, intergalactic scattering. \citet{Koay14} have recently shown that while scatter-broadening in the warm-hot component of the IGM is not expected to be detectable, intergalactic scatter broadening can be significant, $\sim 100\mu$as at 1 GHz and $\sim 3\mu$as at 5 GHz, for sight-lines intersecting within a virial radius of at least one galaxy halo. Targeted, multi-frequency ISS surveys of compact sources behind rich clusters, galaxies or Damped Lyman-Alpha Systems may provide the best chance of detecting extragalactic scatter-broadening through the predicted angular size scaling as $\nu^{-2.2}$, providing a unique probe of the turbulent properties of these regions. Evidence for scatter-broadening in an intervening galaxy has been found from VLBI observations of AGN behind M31 \citep{Morgan13}.

A handful of sources have been found which exhibit unusually rapid variability on timescales $<1$ hour \citep{Kedziora97,Dennett00,Bignall03}, which is modelled as ISS in the weak scattering regime due to very nearby scattering plasma, within tens of parsec of the Sun \citep{Rickett02,Dennett03,Bignall06}. The scattering is determined to be highly anistropic, as has been found also for a number of pulsars displaying diffractive ``scintillation arcs'' in their secondary spectra \citep[][and references therein]{Walker08,Brisken10}.
This suggests scattering in thin ``screens'' in which the electron density fluctuations are subject to extreme magnetic stresses \citep{Tuntsov13}. The specific nature of the material responsible is still unknown.  
Despite two of the three exemplary ``intrahour variables'' being discovered serendipitously, the MASIV Survey found no more examples of such rapid, large-amplitude ISS, implying that the covering factor of nearby screens is very small; $\ll 1$\% of compact sources would be expected to vary on intra-hour timescales. 

The correlation between H-alpha and short-term variability from MASIV suggests that in general, the more typical slower, low-amplitude ISS is probably related to the overall amount of ionized material in the ISM along the line-of-sight, but this is not the case for the intrahour variables. Because AGN angular sizes are typically larger than the Fresnel scale at interstellar distances, they are ``resolved'' by ISS. The smaller the screen distance $L$ from the observer, the larger the angular Fresnel scale, so nearby scattering will tend to dominate the observed variability; not only is the ISS less quenched, but nearby screens also induce faster variations compared with more distant scattering material, since the Fresnel zone crossing time which sets the timescale for weak ISS scales as $\sqrt{L}$ for a given scintillation velocity. ISS behaviour is often observed to be episodic, which can be due to a combination of intermittency in interstellar turbulence and intrinsic changes in source structure, although the relative importance of the two effects remains poorly constrained. This problem can be addressed by comparing longer term source-intrinsic changes in spectral index, flux density, polarization and VLBI structure, with changes in ISS behaviour for a large source sample. 

Finding a large number of rapid intraday variables through SKA Surveys will help to determine the nature of the local scattering plasma in the ISM. While fast intra-day variability can be picked up in undersampled data, extracting useful science would require detailed follow-up observations, either with (a sub-array of) SKA or with other instruments. Well-sampled ISS light-curves across a broad bandwidth, ideally several GHz, can be used to determine the detailed polarization- and frequency-dependent structure of the ultracompact jet with $\mu$as precision. This information can be used to model the magnetic field configuration, geometry, particle density and pressure in the jet \citep{Rickett02,Macquart13}.

\subsubsection{Extreme Scattering Events}~\label{sec:ESEs}
\begin{figure}
\centering 
\includegraphics[angle=270,width=.55\textwidth]{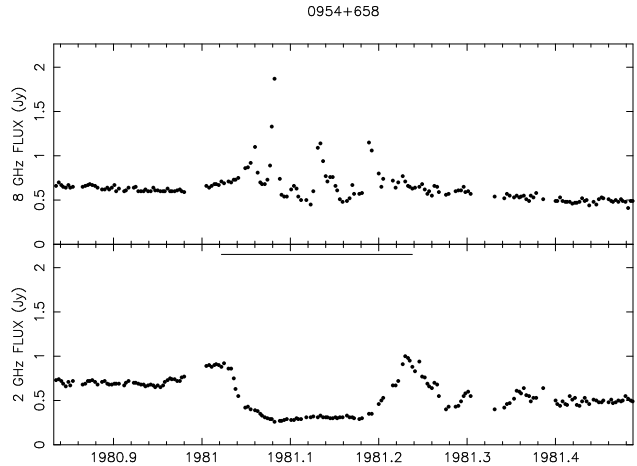}
\caption[]{The extreme scattering event toward QSO~0954$+$658, the
prototypical and exemplar \hbox{ESE}.  The top panel shows the 2~GHz
flux densities measured by the Green Bank Interferometer, and the
bottom panel shows the 8~GHz flux densities. Data are from \citet{rickett06}.}
\label{fig:lightcurve}
\end{figure}

Extreme scattering events (ESE) are a class of dramatic changes in the
flux density of radio sources \citep{fdjws94}.  They are typically
marked by a decrease, as large as $\gtrsim 50$\%, in the flux density near 1~GHz
for a period of several weeks to months, bracketed by substantial
increases, viz.\ Figure~\ref{fig:lightcurve}.  Because of the
simultaneity of the events at different wavelengths, the time scales
of the events, and light travel time arguments, ESEs are likely due to
strong scattering by ionized structures in the Galactic ISM 
\citep[][but see \citealt{ww98}]{fdjh87a,rbc87}.  First
identified in the light curves of extragalactic sources, ESEs have
since been observed during a timing program of the pulsars
PSR~B1937$+$21 \citep{cblbadd93,lrc98} and
PSR~J1643$-$1224 \citep{mlc03}.

While the focus of this chapter is on the variability of \hbox{AGN},
ESEs probably are
related to ``strong fringing events'' observed in dynamic spectra from
pulsars \citep{hill:05}.  In those events, the dynamic spectra---intensity as a function of time and frequency---show ``fringes,''
consistent with the interference from multiple images of a pulsar
resulting from strong lensing along the line of sight to the pulsar.

Modeling of ESE light curves leads to inferred densities $n_e \gtrsim
10^2$~cm${}^{-3}$ within these ionized structures \citep{rbc87,cfl98}.
In turn, these densities imply pressures $nT \sim 10^6$~K~cm${}^{-3}$ or
more, well in excess of the ``average'' interstellar pressure $nT \sim 
3000$~K~cm${}^{-3}$ \citep{kh88}.

Outstanding issues related to ESEs include the following:
\begin{itemize}
\item What is the mechanism by which they occur?  Most models predict
that ESEs arise from strong refraction through relatively isolated
structures.  Observational support for this prediction is mixed.
Pulse times of arrival from the pulsars that have undergone an ESE do
show an increased scatter, but no increase in the pulse width---as
would be expected from the extra path length caused by the refraction.
However, VLBI observations of the source B1741$-$038 while it was
undergoing an ESE show its angular diameter to \emph{increase},
contrary to expectations from a purely refractive model
\citep{lazioetal00}. 
One obvious resolution is to model the ESE
structures as being internally turbulent, but largely refracting
structures. The serendipitous detection of multiple imaging of the quasar 2023+335 observed with VLBI during an ESE was recently reported, and found to be consistent with strong refraction through a turbulent plasma structure \citep{pushkarev:13}.

\item What is the relationship of ESE structures to other phases of
the ISM? or are ESE structures even
interstellar?  ESE structures may represent relatively isolated
structures, perhaps in pressure balance with a lower density, higher
temperature ``background'' phase \citep{ccc88}; they may represent
manifestations of the most extreme pressure ranges within the ISM and
thereby trace energy input into the ISM \citep{jt01}; they may result from a
low level of cosmic-ray ionization within an otherwise neutral tiny
scale atomic structure (\hbox{TSAS}; \citealt{h97}); or they may not be
interstellar at all, but due to photoionized molecular clouds in the
Galactic halo \citep{ww98}.  \ion{H}{i} absorption measurements during
the ESE of B1741$-$038 show no statistically significant change in the
\ion{H}{i} line during the \hbox{ESE} \citep{lazioetal01}.  However,
the limits on the changes in optical depth, and therefore \ion{H}{i}
column density, only marginally rule out a connection between ESE
structures and TSAS in the \hbox{ISM}.  Any change in the \ion{H}{i}
column density is limited to be $\Delta N_H < 10^{18}$~cm${}^{-2}$,
while TSAS has a typical column density of $3 \times
10^{18}$~cm${}^{-2}$.  Geometrical factors (e.g., the line of sight
did not pass through the center of the cloud) could easily account for
this difference.  Conversely, the velocity range of the
\ion{H}{i} absorption measurements was no more than 250~km~s${}^{-1}$.
If ESE structures are halo objects, the allowed velocity range could
approach 1000~km~s${}^{-1}$.  Thus, the existing \ion{H}{i} absorption
observations cannot constrain significantly either of the two
competing models.

\item What is the frequency with which ESEs occur and what is their
sky distribution?  The Green Bank Interferometer (GBI) monitored
approximately 30 sources for over a decade and an additional 120
sources for five years.  A total of~12 ESEs were found in this
monitoring program.  However, a six-month monitoring program with the
NRAO 300~ft telescope found a comparable number of ESEs
(R.~Fiedler~1998, private communication).  The 12 ESEs observed in the
GBI program appear to be located near the edges of radio Loops~I--IV
\citep{fdjws94}, structures thought to be old supernova remnants.  If
this is correct, old supernova remnants could provide the
high-pressure environment in which ESE lenses could survive
\citep{ccc88}.  However, the radio Loops cover a large fraction of the
sky, and the sky distribution might also be consistent with a halo
population of objects.
\end{itemize}

Since the cessation of the GBI monitoring program in~1994
\citep{fiedleretal87b,wfjsfjmm91,waltmanetal99}, there has been no
monitoring program observing a sufficiently large number of sources,
at a low enough radio frequency, and with frequent enough time
sampling to have a realistic chance of detecting the requisite number
of ESEs to address these questions. 
Although some new programs aim to tackle this in the lead-up to SKA Phase 1, e.g.\ the ASKAP VAST project \citep{Murphy13}, these cannot match the efficiency of potential SKA surveys to detect a sufficient number of ESEs to address their statistics (\S\ref{sec-rates}).

Identification of ESEs as they occur is critical in order to determine the nature of the refracting ``lenses'' responsible. Multiple imaging on milliarcsecond scales is predicted at strong caustics, and VLBI monitoring of ESEs will differentiate between competing models for ESEs \citep{pen:12}. Detailed radio follow-up over a broad frequency range, and observations at other wavelengths, including optical/infrared, will also help to rule out or support various models.

%

%% file: subsec-classification.tex
\subsection{Determining the origin of AGN radio variability}
\label{sec:classification}

\begin{figure}
\centering 
\includegraphics[angle=270,width=.6\textwidth]{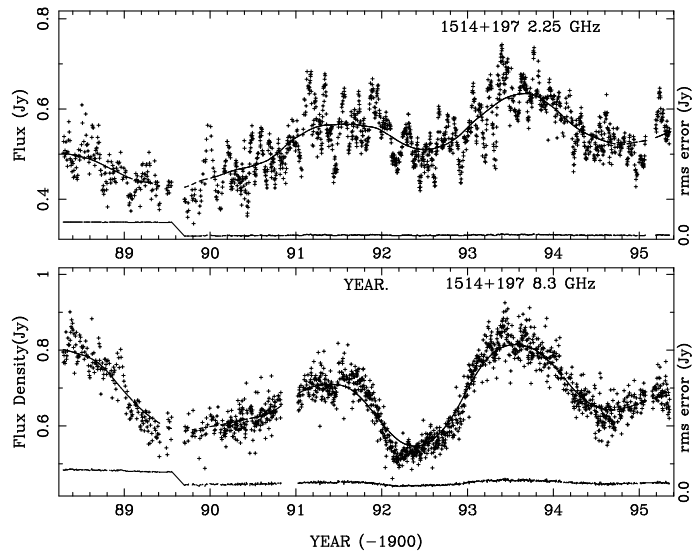}
\caption[]{Flux density at 2 and 8 GHz for 1514+197 from the Green Bank Interferometer monitoring program, analysed by \citet{rickett06} (figure reproduced with permission). The solid curve is the time series (Jy) smoothed over 365 days with a truncated cosine function. It is interpreted to be dominated by intrinsic variation in the synchrotron radiation from the source. The short-term deviations from the line are due to ISS. The trace below each curve gives the rms error in a single measurement (on the same scale) relative to the horizontal axis. This demonstrates the value of long-term, multi-frequency monitoring with high cadence to determine the origin of variability.}
\label{fig:Rickett06}
\end{figure}

Observations at multiple frequencies and classification of light curves based on their structure function, power spectral density, and other statistics will help to disentangle intrinsic and extrinsic causes of variability. \citet{Ofek11} argued that a large fraction of variability at 1.4 GHz must be RISS, based on a flat structure function between $\sim 200-1800$ days for a sample of 43 variables. 
\citet{rickett06}, in an analysis of GBI monitoring data for 146 sources, applied a simple filter on the time series to separate slow intrinsic variations from fast ISS, confirming that both processes are important at the observed frequencies near 2 and 8 GHz (see Figure~\ref{fig:Rickett06}).

 It is expected that propagation effects may dominate at low frequencies, below $\sim 2$~GHz, while large-amplitude variations at higher frequencies, $> 8$~GHz are more likely intrinsic --- at least on timescales of weeks to months and longer, although both ISS \citep{savolainen:08,kara:12} and ESEs \citep{pushkarev:13} have been observed at 15~GHz. However, a test shows no statistical correlation between line-of-sight Galactic H$\alpha$ emission and 15~GHz variability on timescales of months to years from published OVRO data, indicating intrinsic variability dominates, whereas there is a strong correlation between H$\alpha$ emission measure and inter-day variability at 5--8~GHz, indicating that the latter is predominantly ISS. 
At these intermediate frequencies, both intrinsic and ISS-induced variability are likely to be important. Concurrent lightcurves, including polarization information, for large numbers of AGN, using Band 5, 4.6--13.8~GHz, of SKA1-MID---and MeerKAT's X-band receivers---along with two lower frequency bands, will provide a powerful discriminant between intrinsic and extrinsic variability. AGN lightcurves from SKA1-LOW, at 50--300~MHz, and SKA1-SUR, at 650--1670~MHz (PAF Band 2), will likely be dominated by extrinsic variability, although the panchromatic view provided by the wide frequency coverage of SKA, in conjunction with instruments such as the Large Synoptic Survey Telescope (LSST) and high energy facilities, will provide a powerful discriminant of the source of variability for a range of objects.

 The largest amplitude variations due to ISS are usually seen close to the transition between weak and strong scattering, which is dependent on Galactic latitude but typically $\sim 5$ GHz \citep{Kedziora97,Walker98}. When the source is much larger than the angular size of the first Fresnel zone at the scattering screen distance, then weak ISS will be suppressed, but slower refractive ISS may still be observed at lower frequencies in the strong scattering regime. Intrinsic variability at centimetre wavelengths will typically have longer timescales and larger amplitude than scintillation. A preliminary comparison of modulation indices from four years of OVRO data \citep{Richards14} at 15 GHz --- predominantly intrinsic variability --- and short-timescale (2 day) modulation indices from the MASIV Survey at 5 GHz --- predominantly ISS --- for sources common to both samples shows a significant correlation: sources which show the largest amplitude ISS are also more intrinsically variable. This finding, yet to be published, indicates that intrinsic source compactness, core dominance and beaming are important factors for ISS, in addition to the properties of the ISM on a particular line of sight.

ESEs may be distinguished from intrinsic variability or ``normal'' ISS through their strong frequency dependence, the presence of caustics at frequencies of at least a few GHz, and the overall time-symmetric behaviour in the light curves, including a sharp flux density decrease (see Figure~\ref{fig:lightcurve}). As discussed in Section~\ref{sec:ESEs}, identifying ESEs as they occur is critical to address the nature of the lensing structures responsible, by obtaining detailed follow-up observations at high angular resolution and at other wavelengths.

Observations of variability, both intrinsic and due to ISS, may additionally provide a means of distinguishing between emission due to AGN and star formation for large samples of the faint radio source population. The survey speed of SKA will enable high cadence monitoring of sample sizes vastly in excess of what is achievable with existing telescopes.

%% file: sec-SKA-capabilities.tex
\section{Time domain studies of AGN with the SKA}
%

\subsection{Source statistics - expected rates of variability}\label{sec-rates}

The VLA FIRST Survey \citep{Becker95} measured an average source density of $\sim 90$ per square degree down to 1 mJy at 1.4 GHz, the vast majority of which are AGN. Below this flux density, star forming galaxies become more numerous, although a significant fraction of sources are found to be AGN-driven (see e.g.\ Figure 4 of \citealt{Norris11}, and references therein). 
The mJIVE-20 VLBA Survey of FIRST sources \citep{Deller14} 
found an increasing fraction of compact components down to 1 mJy, with 30-35\% of sources being core-dominated and $\sim 10$\% being completely compact. Such compact sources are likely to be variable. \citet{Deller14} obtained a lower limit on the highly variable source fraction at 1.4 GHz based on the number of sources which showed an increase in the VLBI flux density over the VLA FIRST Survey flux density measured some 10--15 years earlier. This number is an underestimate of the true number of variables because the different resolutions of the VLA and the VLBA mean that only a relatively large increase in flux density of core-dominated sources could be measured, since an observed decrease in flux density could be entirely due to resolution effects rather than variability. Based on this, a conservative lower limit to the variable fraction at 1.4 GHz, for $>25$\% variability on timescales of years, is 0.7\% of {\it all} radio sources in the 1--20 mJy range. This is in agreement with results from comparing flux densities for sources detected in multiple epochs of the FIRST survey, where around 0.6\%\ of sources vary by 25\%\ or more \citep{thyagarajan:11}. Down to flux density levels of 40\,$\mu$Jy, \citet{mooley:13} found 1\% of unresolved sources variable at the 4$\sigma$ level on timescales from 1 day to 3 months, with no evidence that the fractional variability changes along with the known transition of radio-source populations below 1 mJy. Optical identifications showed the variable radio emission to be associated with the central regions of an AGN or star-forming galaxy.

The fraction of sources identified as variable naturally increases as source lightcurves are measured with higher precision and over a wider range of timescales and frequencies. Close to the detection threshold of any survey, it is difficult to determine whether a source is variable or not \citep[see, e.g.\ fig.\ 26 of][for a quantitative example for the case of the Allen Telescope Array]{croft:13}. 
A strawman survey with SKA1-SUR or SKA1-MID in band 2, $\sim 1.2$\,GHz, would be capable of surveying the entire visible sky every day to an RMS sensitivity of around 100\,$\mu$Jy beam$^{-1}$, enabling $10\sigma$ detections of variability at the 50\%\ level for sources brighter than $\sim 2$\,mJy. Almost all of these sources will be AGN. Of order $10^6$ sources, similar to the entire catalogs of previous surveys such as FIRST and NVSS, could be monitored for such variability on a daily basis, or larger numbers of fainter sources by binning data. For bright sources, high-precision lightcurves in total as well as polarized flux density will enable measurements of variability with an order of magnitude or more better precision than is possible with existing facilities. The accuracy and stability of calibrated flux density measurements is a critical parameter here; as an example, the MASIV VLA Survey achieved $<1$\% rms stability over each observing session. Because of the large number of available in-beam calibrators, many of which will be steep-spectrum and not significantly variable, in principle SKA instruments should be able to achieve at least this level of precision for flux density measurements, but careful implementation of calibration procedures will be required.

At flux densities $\lesssim 1$\,mJy, the population begins to transition from AGN-dominated to starburst-dominated, although even at 1.4\,GHz flux densities of $\sim 50 \mu$Jy, AGN still make up $\sim 25$\%\ of radio sources \citep{seymour:08}. Starburst galaxies are not expected to be variable at radio wavelengths, with the exception of rare transient events such as radio supernovae, but low-luminosity AGN---including those that are technically classified as ``radio-quiet'' but nevertheless have low-power radio emission---will still be present, both in ordinary galaxies and in some of the starburst galaxies themselves. The high resolution---tens of milliarcseconds---enabled by SKA2 will help to distinguish nuclear point sources from emission due to a surrounding disk, but variability will also be an important clue to the presence of AGN whose emission might otherwise be swamped by the host galaxy. Variability may also be key in identifying and classifying counterparts to sources found by CTA or other high energy facilities, where positional coincidence alone may be insufficient \citep{giroletti}.

\subsection{Frequency coverage}

Intrinsic AGN variability is more extreme towards higher frequencies. Therefore SKA1-MID band 5, covering 4.6--13.8~GHz, along with two lower frequency bands, would be extremely useful for distinguishing intrinsic from extrinsic changes. While not critical for this work, an expansion to 24\,GHz as part of SKA2 would enable monitoring observations at similar frequencies to existing long-term surveys focusing on intrinsic variability (e.g., Fig.~\ref{fig:iiizw2}).

The inclusion of SKA1-MID or SKA1-SUR band 2, covering 1.4~GHz, would allow measurements of \ion{H}{i} absorption towards sources undergoing ESEs (section~\ref{sec:ESEs}).
The optimal frequency range for ESE searches is not yet well determined. Refractive effects are stronger towards lower frequencies, and the ``cross-section'' for events occurring may also be higher due to larger source sizes, however larger source sizes are also a problem as they result in the ESE light curves being ``smeared out''.  ESEs have also been seen at frequencies as high as 15\,GHz \citep{pushkarev:13}. 
Ideally, monitoring would be done over a wide range of frequencies; this could be facilitated by the use of sub-arrays observing simultaneously in different frequency bands.  
The appearance of caustics at frequencies of a few GHz helps to unambiguously identify ESEs (see Figure~\ref{fig:lightcurve}).

For ISS, the largest amplitude variations may be observed close to the transition frequency between weak and strong scattering, typically a few GHz. Weak ISS occurs on shorter timescales than refractive ISS. In general, AGN components are too large to exhibit the narrow-band, short-timescale diffractive ISS observed for pulsars in the strong scattering regime \citep[][but see \citealt{Macquart06}]{Dennison81}.

\subsection{Modeling intrinsic variations}

In the brightest AGN, blazars, the variability is often modeled with shock-in-jet models \citep{marscher85,hughes85}. Recently, detailed modeling of blazar variability during high activity periods was conducted using 4.8, 8 and 15\,GHz total intensity and linear polarization archival data from the University of Michigan Radio Astronomy Observatory \citep{aller14}. The variations were modeled with shocks moving down the jet, and the results showed that no extreme jet conditions are required to form the shocks. The jet and shock parameters obtained were also very similar in all sources modeled. Multiple frequencies providing spectral index information and inclusion of linear polarization were essential for constraining the model parameters. The higher frequencies of SKA1-MID will provide information about the intrinsic variability of a much larger set of AGN, and modeling of a large sample of sources will become possible. In particular, SKA will have unprecedented ability to monitor changes in the weak, typical $V/I < 1$\%, but highly variable circular polarization (CP) component \citep{homan:06}, a powerful probe of the evolution of magnetic fields in relativistic jets \citep{agudo}. CP is associated with the most compact components of AGN, which are also subject to ISS. Unusually strong and intra-day variable CP has been observed for the BL Lac object PKS~1519$-$273, with the variability readily explained as ISS of a component with up to 3.8\% CP, although the origin of the CP itself is unclear \citep{Macquart00}.

The great sensitivity of SKA will enable the extension of variability modeling to lower luminosity sources. Monitoring of broad absorption line (BAL) QSOs and other radio quiet objects, such as Mrk\,231, has already revealed variations similar to blazars. 
Although large scale jets and lobes do not develop in these
objects, likely due to local environmental conditions---perhaps the
presence of the BAL wind, highly energetic flares
can still occur close to the core \citep{reynolds09,reynolds13}.
However, with few examples, it is difficult to establish if the flaring is a common property in so-called radio quiet AGN. With SKA, the number of low luminosity AGN monitored will be extremely large, allowing detailed studies of the variability and comparison to blazars. This will give insights into the jet formation processes and bridge the gap between variations in jetted systems around stellar-mass black holes and the most extreme blazars.

Another interesting group of fainter AGN that have recently received a lot of attention are the Narrow-Line Seyfert I Galaxies (NLS1). Only 7\% of NLS1s are formally radio loud \citep{komossa06}, but interestingly the radio loud ones show many similar properties to blazars, for example, several of them are detected in gamma rays by {\it Fermi} \citep{abdo09}. They also show fast variations in their radio light curves, similar to blazars \citep{foschini11,dammando12}. With SKA we will be able to probe the variability in both radio loud and radio quiet NLS1 objects, shedding light on why some of them are radio loud and others not. If no significant variability is detected, it implies that the sources are not Doppler beamed and are viewed at larger angles to the line of sight.

In general, the large number of AGN observed by SKA will allow statistical analysis of AGN variability over large populations. By calculating the power spectral densities 
of the light curves, we can gain understanding of the nature of the variations. In cases where significant polarization is detected, we obtain information about the underlying magnetic field structure in these objects. For example, if the polarization angle is stable over a long time period, it implies that large-scale ordered magnetic fields play a role \citep{aller03}. The strength of SKA will be the ability to study the variations in a large number of sources over a large luminosity range, shedding light on AGN unification.

\subsection{Use case: synoptic surveys with outer antennas of SKA1-MID}

It is anticipated that a number of SKA1-MID projects will not make use of the longest available baselines. Using the outer antennas to conduct a monitoring programme while the inner antennas conduct other observational programmes will be an extremely valuable operational mode. The outer antennas could be used for monitoring sufficiently bright targets in order to study their ISS and intrinsic variability, as well as searching for ESEs (and other transient events).
Flexible sub-arraying to allow simultaneous observations at multiple frequencies, and/or of multiple targets, would be an important capability. The value of sub-arraying for time domain work was successfully demonstrated by the MASIV Survey, which used the VLA split into 5 subarrays to monitor $\sim 700$ fields with two-hourly cadence \citep{Lovell03}.

\subsection{Capabilities of SKA1 with 50\% reduced sensitivity}

The early science phase of SKA deployment, although not reaching full SKA1 sensitivity, nevertheless represents a huge leap over existing surveys, and a transition between those possible with MWA, MeerKAT, and ASKAP, and the full SKA1. 
The main effect of 50\%\ reduced sensitivity relative to SKA1 on the surveys discussed here would be a decrease of a factor $\sim 3$ in the number of sources that could be monitored to a given flux density precision. For programmes where sub-arraying is the desired observing mode, the early science phase would simply use fewer sub-arrays. Limited instantaneous $(u,v)$ coverage of the early roll-out phase may pose problems for the type of snap-shot monitoring observations discussed here; in any case an accurate sky model to sufficient depth will be required to search for variability. 

\subsection{Prospects for SKA2}

Current VLBI surveys reach milliarcsecond resolutions, corresponding to parsec scales in AGN. Resolution of the baseline SKA-MID 
is an order of magnitude or so poorer. Including sufficient sensitivity on long baselines as part of SKA2 \citep[see][]{Paragi14} would enable the expansion of existing monitoring programs to much larger numbers of sources. Polarization capabilities will be particularly important in enabling identification and tracking of jet components, characterizing changes in magnetic fields and their connection to flares, and shedding light on jet composition and the interaction of jets with the ambient medium. Simultaneous monitoring of large numbers of sources in conjunction with high-energy all-sky instruments will enable us to constrain the dissipation of energy during rare AGN outbursts.

One of the limitations of AGN monitoring at high frequencies is the decrease of survey speed with the square of the frequency, assuming single pixel feeds, constant sensitivity, and flat spectrum sources. A factor 20 increase in field of view with SKA2-MID would enable survey speeds at 6\,GHz comparable to those with SKA1-MID at 1.4\,GHz, 
enabling the kind of surveys discussed above to be undertaken at these higher frequencies, where intrinsic variability becomes increasingly important.